# TERAHERTZ DETECTION WITH δ-DOPED GaAs/AlAs MULTIPLE QUANTUM WELLS

D. SELIUTA[a], B. ČECHAVIČIUS[a], J. KAVALIAUSKAS[a], G. KRIVAITĖ[a], I. GRIGELIONIS[a], S. BALAKAUSKAS[a], G. VALUŠIS[a,*], B. SHERLIKER[b], .M. P. HALSALL[b], M. LACHAB[c], S. P. KHANNA[c], P. HARRISON[c], AND E. H. LINFIELD[c]

[a] Semiconductor Physics Institute, A. Goštauto 11, LT-01108 Vilnius, Lithuania

[b] School of Electronic and Electrical Engineering, University of Manchester Manchester, United Kingdom

[c] School of Electronic and Electrical Engineering, University of Leeds, Leeds LS2 9JT, United Kingdom

The authors demonstrate selective detection of terahertz radiation employing beryllium δ-doped GaAs/AlAs multiple quantum wells. The sensitivity up to 1 V/W within 4.2−7.3 THz range at liquid helium temperatures is reached. The Franz-Keldysh oscillations observed in photo- and electro-reflectance spectra allowed one to estimate built-in electric fields in the structures studied. It was found that the electric field strength in the cap layer region could vary from 10 kV/cm up to 26 kV/cm, depending on the structure design and temperature.

## 1. Introduction

The large range of potential applications for terahertz (THz) radiation has given an increased impetus to the search for new concepts in the development of compact THz sources and detectors. Quantum structures, fabricated by molecular beam epitaxy (MBE), are an attractive choice as they allow, depending on the precise structure design, different physical mechanisms to be used for the emission/detection of THz radiation. An excellent example of a compact emitter is the THz quantum cascade laser [1], whilst in detection, THz QWIPs (quantum well infrared detectors [2]) or HEIWIPs (heterojunction interfacial work function internal photoemission detectors [3]) are being developed. These devices exploit *vertical carrier transport* which can be engineered by varying the widths of the wells/barriers and the doping profile. Alternatively, one can rely on the properties of carriers *along the structure layers*. This is well illustrated for detection of THz radiation by the study of the excitation of plasma waves [4] or non-uniform electron heating in a two-dimensional electron gas [5].

* corresponding author; e-mail: valusis@pfi.lt

In this letter, a comprehensive experimental − terahertz photocurrent and modulated reflectance spectroscopy in optical range − study of *p*-type (beryllium) δ-doped GaAs/AlAs multiple quantum wells (MQWs) as possible selective sensors for the THz range is presented.

## 2. Samples and experimental techniques

The MQW structures of various designs were grown on semi-insulating GaAs substrates by MBE. The samples were δ-doped at the well centre with Be acceptor atoms, the AlAs barrier was kept equal to 5 nm. The main parameters of the studied MQWs are given in the Table I below.

**Table I.** Characteristics of the studied MQWs: the repeated period, the quantum well width ($L_w$) and the δ-doping Be density $N_A$.

| Samples | Periods | $L_w$ (nm) | $N_A$ (cm$^{-2}$) |
|---|---|---|---|
| L151 | 60 | 15 | $5\times10^{12}$ |
| L152 | 60 | 15 | $5\times10^{11}$ |
| 1392 | 40 | 20 | $2.5\times10^{12}$ |
| 1794 | 200 | 10 | $5\times10^{10}$ |

The MQW structures were studied by measuring the THz photocurrent using, as the THz source, either a free electron, or an optically-pumped molecular, THz lasers operating in a pulse mode. To determine the distribution and values of the built-in electric fields in the structures, Franz-Keldysh oscillations (FKOs) in the photoreflectance (PR) and contacless electroreflectance (CER) spectra were recorded. For the PR studies, a He-Ne (632.8 nm) or LED (470 nm) light beams as the excitation sources were employed. In the CER experiment, a condenser-like system was utilized. An ac modulation voltage in the range of 300 – 500 V was applied to top transparent electrode to provide modulating field.

The experiments were performed within the 4.2–300 K temperature range employing closed-cycle helium cryogenic system.

## 3. Experimental results and discussion

Before proceeding to the THz photocurrent, we have determined built-in electric fields in the MQW structures from the study of the PR/CER spectra shown in Fig. 1a. As one can see, for moderately and highly doped samples the spectra are dominated by the bulk-like oscillating signal arising in the vicinity of the GaAs fundamental gap $E_g$ (1.424 eV at 300 K and 1.505 eV at 90 K). These PR/CER features we associated with FKOs indicating the existence of an internal electric field in the samples. By analysing FKOs above the GaAs band edge, it is possible to calculate the built-in electric field following the procedure given, for instance, in Ref. 6.

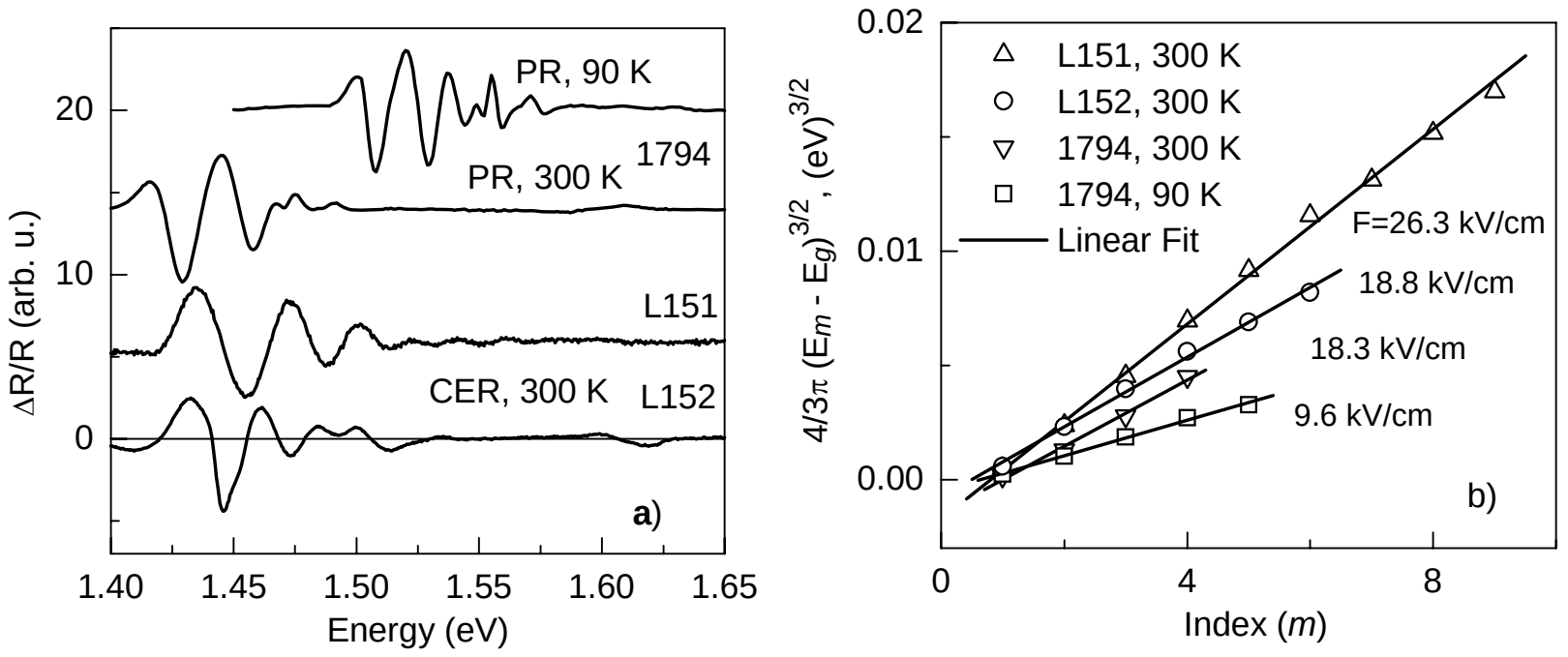

**Fig. 1a** PR and CER spectra of various MQW samples measured at 300 and 90 K temperatures. The FKOs are clearly resolved above the GaAs band edge. **Fig. 1b** Plots of the quantity $(4/3\pi)(E_n\text{-}E_g)^{3/2}$ versus index *m* of the FKOs for the Be- doped GaAs/AlAs MQW structures.

This analysis and, plus, surface photovoltage spectroscopy data (not given here) allowed to determine the built-in electric field strength in the cap layer region. Their values are within 10 −26 kV/cm, depending on the structure design and temperature (Fig. 1b).

Terahertz photocurrent data are presented in Fig. 2a. The origin of the signal is *photothermal ionization* of the Be acceptors, which binding energy is QWs width-dependent. Therefore, the detection frequency can be tuned across the THz range by variation of the impurity confining potential. It is worth noting two distinctive features: *first,* signal appears below the ionization energy (under 4 THz) is related to excited acceptor states $2P_{3/2}$ and $2P_{5/2}$, *second,* the signal decreases with the increase of doping ($\geq 10^{12}$ $cm^{-2}$) due to the approach of the Mott transition [7]. Tunability of free electron laser enabled to observe a strong absorption line at a frequency of around 5.4 THz owing to the *intra-acceptor absorption* of the bound holes (Fig. 2b).

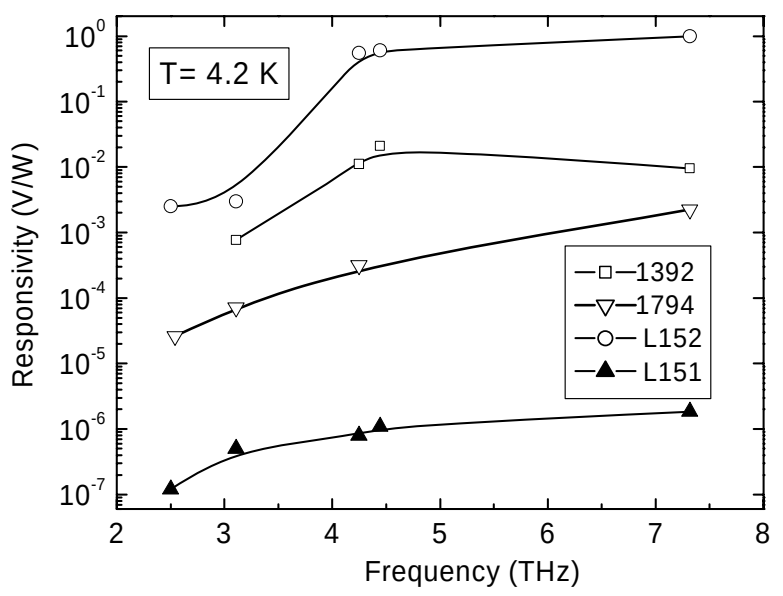

**Fig. 2a**. THz photocurrent spectra in various QWs measured using optically pumped molecular THz laser.

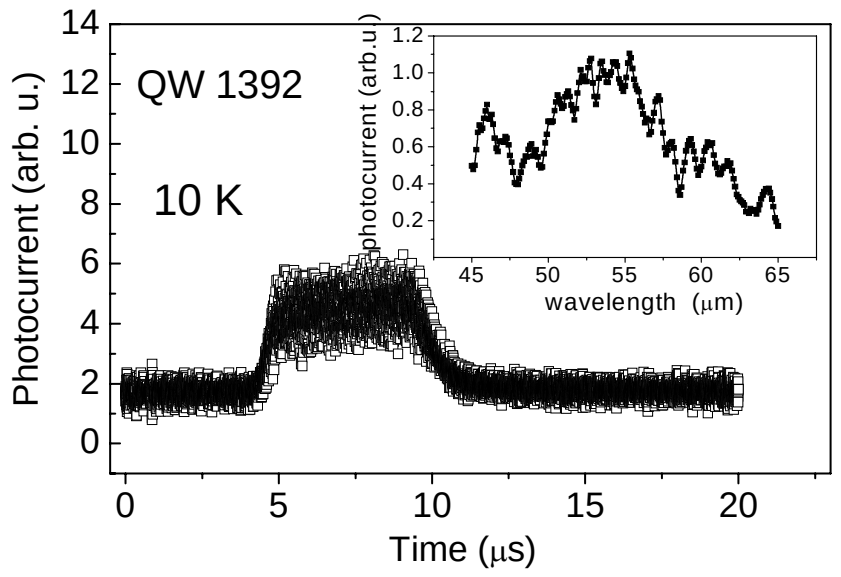

**Fig. 2b**. Transient of THz photocurrent in QW 1392 recorded around 5.4 THz (55 μm) using free electron laser. Insert – spectrum of the THz photocurrent.

## Acknowledgements

The work was supported, in part, by the Lithuanian State Science and Studies Foundation under contract C-07004.